\documentstyle{article}
\title{Effective action  in a quantized metric}
\author{ Z. Haba\\Institute of Theoretical Physics, University of Wroclaw,
\\50-204 Wroclaw, Plac Maxa Borna 9,
Poland\\e-mail:zhab@ift.uni.wroc.pl}
\date{}
\begin{document}
\maketitle
\begin{abstract}
 We calculate the effective action in Yang-Mills  and scalar $\phi^{4}$ quantum field
 theory with quantized scale invariant metric treated
 nonperturbatively
in $d=4$ dimensions. There is no charge renormalization
in the one-loop order for matter fields. We show that
the electromagnetic energy of point
charges can be finite. The temperature dependence of the effective
action in inflationary models is changed substantially as
a result of an interaction with quantum gravity.

\end{abstract}
\section{Introduction}
The effective action is derived as a result of an integration
over quantum fields with some classical sources. At the same time
it can be given a meaning of the classical action whose tree
approximation takes into account  the quantum effects. It gives a
useful illustration of the quantum corrections to classical
phenomena. In this paper we show that the form of the effective
action for fields interacting with gravity follows from the
scaling properties of quantum gravity. We assume that quantum
gravity is scale invariant at short distances with the scale
dimension $\gamma$. When expressed in terms of the metric $g$
this means that  $\lambda^{2\gamma}g^{\mu\nu}(\lambda x )$ and
$g^{\mu\nu}(x)$ have the same correlation functions. We calculate
the 1-loop generating functional for matter fields in an external
gravitational field. Then, the (quantum) correlation functions of
the gravitational field are assumed to be scale invariant. An
approximate evaluation of an average of the generating functional
over the gravitational field leads (after the Legendre
transformation) to an effective action which is substantially
different from the perturbative one. The inverse propagator grows
faster in momenta than the free one. Moreover, the 1-loop
corrections to the effective action do not receive the
logarithmic corrections characteristic to the renormalizable
quantum field theories.

 \section{The scalar heat kernel}
We repeat some  steps of our ealier paper \cite{habalet}
where an interaction with a scale invariant gravity has been  discussed.
We express an average of the heat kernel of the Laplace-Beltrami operator
$2{\cal A}$ over the metric $g$ by means of the path integral
\begin{equation}
\begin{array}{l}
\langle K_{\tau}(x,y)\rangle=\langle(\exp\tau{\cal A})(x,y)\rangle=
\int {\cal D}g\exp\left(-\int dx {\cal
L}_{g}\left(g\left(x\right)\right)\right)
\cr
\int {\cal D}q\exp(-\frac{1}{2}\int g^{\mu\nu}\frac{dq_{\mu}}{dt}
  \frac{dq_{\nu}}{dt})
 \delta\left(q\left(0\right)-x\right)
  \delta\left(q\left(\tau\right)-y\right)
  \end{array}
  \end{equation}
where ${\cal L}_{g}$ is the gravitational Lagrangian.
 In order to obtain the behavior of $\langle K_{\tau}(x,y)\rangle $
 for a small $\tau$ it is  sufficient to assume the scale invariance
 of the metric $g$ at  short distances. We can also derive this
 short time  asymptotics directly from the functional integral
 (1) assuming a  scale invariance of the action
 \begin{equation}
 \int dx {\cal L}_{g}\left(\lambda^{2\gamma}g\left(\lambda
 x\right)\right) =\int dx {\cal L}_{g}\left(g\left(x\right)\right)
 \end{equation}
   In fact,  introducing in eq.(1) a new functional integration
   variable $\tilde{g}=  \lambda^{2\gamma}g\left(\lambda
 x\right)              $ and a new path  $\tilde{q}$
 defined on an interval $[0,1]$ by
 \begin{equation}
q(\tau s)-x= \tau^{\sigma}(\tilde{q}(s)-x)
\end{equation}
where
 \begin{displaymath}
 \sigma=\frac{1}{2}(1+\gamma)^{-1}
 \end{displaymath}
 we can explicitly extract the $\tau$-dependence of
 $\langle K_{\tau}(x,y)\rangle $
 \begin{equation}
 \langle K_{\tau}(x,y)\rangle=\langle E[\delta\left(y-x-
 \tau^{\frac{1}{2}-\sigma\gamma}\tilde{q}\left(1\right)\right)]\rangle
 \end{equation}
 where $E[.]$ denotes an expectation value over the paths
 $\tilde{q}$. Then, an average over the propagator
 (in the proper time representation) behaves as
 \begin{equation}
  \langle {\cal A}^{-1}(x,y) \rangle =\int_{0}^{\infty}
  d\tau \langle K_{\tau}(x,y)\rangle\simeq \vert x-y\vert^{-2 +2\gamma}
  \end{equation}
  For the effective action we need only the diagonal of the heat
  kernel. From eq.(4)
\begin{equation}
\begin{array}{l}
\langle K_{\tau}(x,x)\rangle=\tau^{-\frac{2}{1+\gamma} } \langle
E\left[\delta\left(\tilde{q}\left(1\right) \right)\right] \rangle  \equiv
\tau^{-\frac{2}{1+\gamma} } v(x)
\end{array}
\end{equation}
where the mean value $v$ depends only on x.

\section{The electromagnetic heat kernel}
For a calculation of the effective action we need
the heat kernel of an operator
arising from the electromagnetic Lagrangian
\begin{equation}
W =\frac{1}{4}\int
d^{4}x\sqrt{g}g^{\mu\nu}g^{\sigma\rho}F_{\mu\sigma} F_{\nu\rho}
\end{equation}
We choose the Feynman gauge which results from  an addition to the action
of the term
\begin{equation}
W_{0}=\frac{1}{2}\int d^{4}x
(g^{\mu\nu}\partial_{\nu}(\sqrt{g}A_{\mu}))^{2}
\end{equation}
We write $\tilde{W}=W+W_{0}$ as
\begin{equation}
\tilde{W} =\frac{1}{2}\int d^{4}x\hat{A}_{a}(
g^{\mu\nu}(-\partial_{\mu}\delta_{ac}+\omega_{\mu a}^{c})
(\partial_{\nu}\delta_{cb}+\omega_{\nu c}^{b}) +R_{ab})\hat{A}_{b}
\equiv \frac{1}{2}\int \hat{A}\triangle_{EM}\hat{A}
\end{equation}
where $g^{\mu\nu}=e^{\mu}_{a}e^{\nu}_{a}$,$\hat{A}_{a}=e^{\mu}_{a}A_{\mu}$ , $R_{ab}$ is the Ricci
tensor and the spin connection $\omega$ is just a transformation
of the Christoffel  symbol to the fixed frame
\begin{displaymath}
\omega_{\mu c}^{a}=e^{\nu}_{c}\Gamma_{\mu\nu}^{\sigma}l^{a}_{\sigma }+
l^{a}_{\nu}\partial_{\mu} e^{\nu}_{c}
\end{displaymath}
where $l$ is the inverse matrix to  $e$.
 We are interested in the heat equation
\begin{equation}
\partial_{\tau}\hat{A}=\frac{1}{2}\triangle_{EM}\hat{A}
\end{equation}
The solution can be expressed in the form \cite{ikeda}
\begin{equation}
\hat{A}(\tau,x)\equiv (T_{\tau}\hat{A})(x)=E[{\cal
T}(\tau)\hat{A}(q_{\tau}(x))]
\end{equation}
where ${\cal T}(\tau)$ is a solution of the equation
\begin{equation}
d {\cal T}_{ac}=\omega_{\mu a}^{b}{\cal T}_{bc}
dq^{\mu}+\frac{1}{2}R_{ab}{\cal T}_{bc}d\tau
\end{equation}
From eq.(11) (cp. with eq.(6))
\begin{equation}
\langle K_{\tau}(x,x)\rangle=
\langle E[{\cal T}(\tau)\delta\left (q_{\tau}\right)
]\rangle
\end{equation}
We can apply now the scaling of sec.2 in order to conclude
that
\begin{equation}
\langle K_{\tau}(x,x)\rangle= \tau^{-\frac{2}{1+\gamma}}v(\tau,x)
\end{equation}
where  $ v(\tau,x)$ is
equal to (with the notation as in eq.(6))
\begin{displaymath}
v(\tau,x)=\langle E[{\cal T}(\tau)\delta\left(\tilde{q}\left(1\right)
\right)]\rangle
\end{displaymath}
It is a regular function of $\tau$.
\section{The effective action for the $\Phi^{4}$ scalar field}
We wish to calculate the
generating functional (where $A$ may denote either the gauge
field or the scalar field)
\begin{equation}
Z[J,\Theta]=\int {\cal D}A{\cal D}g\exp\Big(
- \frac{1}{\hbar}(L +JA+g\Theta)\Big)
\end{equation}
In the conventional approach we make a shift $A\rightarrow A+B$ and $g\rightarrow g+h$. We
choose $B$ and $h$ as solutions of classical equations. Then, the
linear terms in $A$ and $ g$ vanish. The integration over the
quadratic term in $A$  gives a determinant. We restrict ourselves
to the effective action for the matter fields in
the 1-loop approximation. Then, $\Theta=0$
and $h=0$. The standard loop expansion is an expansion in the
Planck constant $\hbar$. We assume that $\hbar$ is small but the
gravitational coupling $\kappa$ is large and of the order
$\frac{1}{\hbar}$. This assumption is a mathematical trick which
allows us to treat gravity beyond the one loop in order to see
its non-perturbative effects. It can however have a physical
interpretation. The parameters $\kappa$ and $\hbar$ have a
dimension. Hence, large and small depends on the physical
context. A small length $\sim \sqrt{\hbar}$
(with small $\hbar$ and large $\kappa$ )means that it is small
in comparison to the Planck length $\sim\sqrt{\kappa\hbar}$.

We discuss first the scalar field Lagrangian (it is
technically  simpler and may be relvant for
inflationary models). Let \begin{displaymath}
L=\frac{1}{2}g^{\mu\nu}\partial_{\mu}\Phi\partial_{\nu}\Phi-
\frac{\alpha}{6} \Phi^{4} \end{displaymath}
 So, taking only
diagrams with one scalar loop we obtain
\begin{equation}
Z[J]=\langle \exp\Big(-\frac{1}{\hbar}W(\phi_{c})\Big)\det{\cal M}^{-\frac{1}{2}}\rangle
\end{equation}
where
\begin{equation}
{  \cal    M}=-{\cal A} +\alpha\phi_{c}^{2}
\end{equation}
and $\phi_{c}$ is the solution of the classical equation
\begin{equation}
-{\cal A}\phi_{c}+\frac{1}{3}\alpha\phi_{c}^{3}=\frac{1}{2}J
\end{equation}
$2{\cal A}$   is the Laplace-Beltrami operator. $W(\phi_{c})$
denotes the classical action.  Clearly, it is not simple to
calculate the average in eq.(16) exactly. We can apply the
cummulant expansion
\begin{displaymath}
\langle \exp U\rangle=\exp\Big(\langle U\rangle+
\frac{1}{2}\langle (U-\langle U\rangle)(U-\langle U\rangle)\rangle+...\Big)
\end{displaymath}
The first order approximation in eq.(16) is
\begin{equation}
Z[J]=\exp\Big(-\frac{1}{\hbar}\langle W(\phi_{c})\rangle
-\frac{1}{2}\langle Tr\ln{\cal M}\rangle\Big)
\end{equation}
We write
\begin{equation}
W_{q}=\frac{1}{2}\langle Tr\ln {\cal M}\rangle= -\frac{1}{2}\int
dx\langle\int_{0}^{\infty} \frac{d\tau}{\tau}(\exp\tau{\cal
M})(x,x) \rangle
 \end{equation}
Applying eq.(3) and (20) we obtain
 \begin{equation}
 W_{q}=-\frac{1}{2} \int dx\int_{0}^{\infty} \frac{d\tau}{\tau}
   \tau^{-\frac{2}{1+\gamma}} \langle E[\delta\left(
\tilde{q}\left(1\right)\right)
   \exp\left(-\tau\int_{0}^{1}ds \alpha\phi_{c}^{2}\left(x+
   \tau^{\sigma}\left(\tilde{q}_{s}-x\right)\right)\right)]\rangle
\end{equation}
$W_{q}$ as written is infinite. In order to define
a finite expression we make the $\zeta$-function regularization
replacing $\frac{d\tau}{\tau}$  by $\frac{d\tau}{\tau^{1-z}}$.
Then, $\frac{d}{dz}{\cal M}^{-z}_{\vert z=0}$
which can be considered as a definition of $ Tr\ln{\cal M}$
 is well-defined .
Up to a finite renormalization the $\zeta$-function definition
is equivalent to a definition by a counterterm subtraction   \cite{schwarz}
 \begin{equation}
 \begin{array}{l}
 W_{q}=-\frac{1}{2} \int dx\int_{0}^{\infty} \frac{d\tau}{\tau}
   \tau^{-\frac{2}{1+\gamma}} \langle E[
   \delta\left(
\tilde{q}\left(1\right)\right)
\Big(
   \exp\left(-\tau\int_{0}^{1}ds \alpha\phi_{c}^{2}\left(x+
   \tau^{\sigma}\left(\tilde{q}_{s}-x\right)\right)\right)
   \cr
   -1
       +\tau\alpha\phi_{c}^{2}(x)\Big)]\rangle
       \end{array}
\end{equation}
The behavior for a large $\phi$ depends on a small $\tau$. Then,
we can use the approximation $\tilde{q}_{s}-x\approx 0$ in eq.(22).
In such a case

 \begin{equation}
 \begin{array}{l}
 W_{q}\simeq - \int dx \langle E[
    \delta\left(
\tilde{q}\left(1\right)\right)
]\int_{0}^{\infty} \frac{d\tau}{\tau}
   \tau^{-\frac{2}{1+\gamma}}
 \Big( \exp(-\tau \alpha\phi_{c}^{2}(x)  )
    -1   +\tau\alpha\phi_{c}^{2}(x) \Big) \rangle
     \cr
     \simeq
       \int dx  \langle
     v(x)
       \vert\phi_{c}\vert^{\frac{4}{1+\gamma}}\rangle
\end{array}
\end{equation}
Note that with $\gamma>0$ the subtraction of the term
$\phi_{c}^{4}(x)$ in eq.(22) is unnecessary (there is no coupling
constant renormalization) .
 The generating functional (23) has the same form as the
 one resulting from a $\Phi^{4}$
 theory in $d=4/(1 +\gamma)$ dimensions. In principle, we can solve
 eq.(18) as a power series in $J$ and calculate the
 expectation value over the gravitational field again
 in a power series in $J$. In this way we obtain the
 generating functional $Z[J]$ for the correlation functions
 $\langle \Phi(x_{1}).....\Phi(x_{n})\rangle $.

It is interesting to do it in the zeroth order in $\alpha$.
Then,
\begin{equation}
Z[J]=\langle\exp(\frac{1}{4}\int J{\cal A}^{-1}J)\rangle \simeq
\exp(\frac{1}{4}\int J\langle{\cal A}^{-1}\rangle J)
\end{equation}
The r.h.s. of eq.(24) will give the correct two-point function
for the scalar field in a quantum gravitational field but
only an approximate formula  for $n$-point correlation functions
(although the short distance behavior will be the same).

The effective action $\Gamma$ is defined as
\begin{displaymath}
\Gamma(\phi)=\ln Z[J]-J\phi
\end{displaymath}
where
\begin{equation}
\phi(x)=\frac{\delta\ln Z[J]}{\delta J(x)}
\end{equation}
Hence, in order to obtain the effective action from the
 generating functional $Z[J]$ we need to express $J$ by $\phi$
 from eq.(25). In the zeroth order in $\alpha$
 with the approximation (24) we obtain
 \begin{displaymath}
 \Gamma(\phi)=\frac{1}{4}\int \phi(\langle {\cal A}^{-1}\rangle)^{-1}\phi
 \end{displaymath}
 where   $(\langle {\cal A}^{-1}\rangle)^{-1} $ denotes
 the kernel of an operator inverse to the operator
 determined by the kernel $ \langle {\cal A}^{-1}(x,y)\rangle$
(in momentum space $\langle {\cal A}^{-1}\rangle(k)\sim k^{2+2\gamma}$). When, $\alpha>0$ the inversion is more complicated
   even with the approximation (19). We have to express
 $\langle W(\phi_{c})\rangle $ by $J$.
 In perturbation theory this is a sum of terms  of the form
 \begin{displaymath}
 \begin{array}{l}
 \langle W(\phi_{c})\rangle=\sum\int dx_{1}....dx_{2n}
 \langle  {\cal A}^{-1}(x_{1},x_{2})....{\cal A}^{-1}(x_{2n-1},
 x_{2n})\rangle\prod_{k,l}\delta(x_{k}-x_{l})
 \cr
 J(x_{1})....J(x_{r})
 \end{array}
 \end{displaymath}
 where the number of $\delta$-functions and $r$ are adjusted so that
 we obtain $2n$ coordinates altogether.
 Then, we have to take the
 Legendre transform  (25) of         $\langle W(\phi_{c})\rangle $
 in order to obtain $\Gamma_{tree}(\phi)$.
 The difference between $W(\phi)$ and $\Gamma_{tree}(\phi)$
 is in the replacement of ${\cal A}^{-1}(x_{1},x_{2})...
 {\cal A}^{-1}(x_{2n-1},x_{2n})$ by
 $\langle {\cal A}^{-1}(x_{1},x_{2})...
 {\cal A}^{-1}(x_{2n-1},x_{2n})\rangle $ (and
  an inversion of the averaged kernels).
  For the 1-loop term
 we make the following approximations
 \begin{displaymath}
 \langle W_{q}(\phi_{c})\rangle \simeq W_{q}(\langle
 \phi_{c}\rangle)\simeq W_{q}(\frac{1}{2} \langle {\cal A}^{-1}\rangle J)
 \simeq W_{q}(\phi)
 \end{displaymath}
 where the last step follows from eq.(25)   .
  Hence, with these approximations

  \begin{displaymath}
 \Gamma(\phi)=\Gamma_{tree}(\phi)+\hbar W_{q}(\phi)
 \end{displaymath}
 In the  inflationary models \cite{linde} we need the effective
potential at finite temperature. We consider a time-independent
(three-dimensional) perturbation of the classical expanding
metric. In the comoving frame (moving with the speed of the
expansion) we may treat the resulting metric as a  static metric on a
threedimensional manifold. A
calculation of the effective potential for a static metric is a
straightforward generalization of the one in eq.(20). We have
just to replace $(\exp\tau{\cal M})(x,x)$ by  a sum over integer
$n$ of $(\exp\tau{\cal M})(x_{0}+n\beta,{\bf x};x_{0},{\bf x})$
(periodic boundary conditions in time, here $\beta^{-1}=KT$,
where $T$ is the temperature and $K$ is the
Boltzman constant). The
part involving $x_{0} $ separates and gives just a factor
$(2\pi\tau)^{-\frac{1}{2}}\exp(-\frac{n^{2}\beta^{2}}{2\tau})$.
The sum over $n$ leads to a formula resembling the standard one (
the summation method and the result are similar
to the standard flat case \cite{dolan})
\begin{equation}
W_{q,\beta}(\phi)\equiv W_{q,\infty}+W^{(1)}_{q,\beta}=
W_{q,\infty}+\beta^{-1}Tr\ln(1-\exp(-\beta M))
\end{equation}
where  $W_{q,\infty}\equiv W_{q} $ is defined in eq.(22) and
\begin{displaymath}
M^{2}=-\frac{1}{2}\triangle_{3}+\alpha\phi^{2}= -\frac{1}{2}g^{kl}\partial_{k}\partial_{l}
-\frac{1}{2}\Gamma^{l}\partial_{l}+\alpha\phi^{2}
\end{displaymath}
 $\triangle_{3}$ denotes the threedimensional Laplace-Beltrami
operator and $\Gamma$ is the Christoffel symbol. We must take an
average over the quantum gravitational field in eq.(26). We need
some approximations in order to perform this difficult task. We
set
\begin{equation}
\langle Tr\ln(1-\exp(-\beta M)) \rangle\simeq
 Tr\ln(1-\langle\exp(-\beta M) \rangle)
 \end{equation}
 With this approximation a calculation of the effective
 potential at finite temperature is already simple.
 We may use the formula
 \begin{displaymath}
 \begin{array}{l}
\langle\exp(-\beta M)\rangle=\frac{1}{\pi }\langle\int
dp\exp(ip\beta)
 \cr\int_{0}^{\infty}\int_{0}^{\infty}ds_{1}ds_{2}s_{1}^{-\frac{3}{2}}
 \exp(-s_{2}p^{2})\Big(
 \exp(-s_{1}M^{2})-\exp(-s_{2}M^{2})\Big)\rangle
 \end{array}
 \end{displaymath}
As follows from eq.(6)  the diagonal part of
 the heat kernel in quantum gravitational
 field in $d$ dimensions is the same as the one without
 the gravitational field but in $d(1+\gamma)^{-1}$ dimensions.
 Applying the Dolan-Jackiw formula \cite{dolan} to $3(1+\gamma)^{-1}$
 dimensions we obtain the following integral representation of the
 1-loop effective action
 \begin{equation}
 W^{(1)}_{q,\beta}=2\beta^{-\frac{3}{1+\gamma}-1}\int_{0}^{\infty}du u^{
 \frac{3}{1+\gamma}-1}
 \ln\Big(  1-\exp(-\sqrt{u^{2}+\beta^{2}\phi^{2}})\Big)
\end{equation}
It follows from eq.(28) that there is no logarithmic correction
to the effective action characteristic of renormalizable models.
As can be seen from eq.(28) the expansion for high temperature
 (and small values of the
field) starts with the term
\begin{displaymath}
\beta^{-\frac{2-\gamma}{1+\gamma}}\phi^{2}
\end{displaymath}
The temperature dependence of the phase
transition in inflationary models can be derived either from
this expansion or from the formula for the correlation function
at high and intermediate temperature
\begin{displaymath}
\begin{array}{l}
\langle \Phi(t,{\bf x})\Phi(t,{\bf y})\rangle\simeq
\frac{1}{2}\Big(M^{-1}\exp(-\beta M)\Big)
({\bf x},{\bf y})
\cr
=\frac{1}{2\pi}\int_{0}^{\infty} d\tau \int dp\Big( \exp(-\frac{\tau}{2}p^{2}
-\frac{\tau}{2}M^{2})\Big)({\bf x},{\bf y}) \exp(ip\beta)
\end{array}
\end{displaymath}
To the r.h.s. we can apply the method of sec.2 which leads to the formula
\begin{displaymath}
\langle \Phi(t,{\bf x})^{2} \rangle\sim \beta^{-\frac{2-\gamma}{1+\gamma}}
\end{displaymath}
Let us note that also the temperature dependence of the energy-momentum
tensor is modified by quantum gravity
\begin{displaymath}
\langle T_{\mu\nu} \rangle \sim \beta^{-\frac{4-\gamma}{1+\gamma}}
\end{displaymath}
 The Planck
spectrum of particles at temperature $T$ will be deformed to
\begin{displaymath}
\rho(E)\simeq E^{\frac{3}{1+\gamma}}(\exp(\beta E)-1)^{-1}
\end{displaymath}
We can conclude that the temperature-dependence
of the phase transition in inflationary models will be modified by an
inclusion of quantum gravity .
\section{The effective action for gauge fields}
 We consider next
the non-Abelian gauge theory at zero temperature. With the same
approximations as for the scalar field we obtain
\begin{equation}
Z[J]=\exp\Big(-\frac{1}{\hbar}\langle W(A_{cl})\rangle
-\frac{1}{2}\langle Tr\ln{\cal M}\rangle+\langle Tr\ln {\cal M}_{FP}\rangle\Big)
\end{equation}
where $A_{cl}$ is a solution of the Yang-Mills equations
with an external source $J$ (we shall omit the index $cl$
furtheron), $\det {\cal M}_{FP}$ is the Fadeev-Popov determinant
(Feynman gauge) and
\begin{equation}
{\cal M}_{\mu\nu}^{ab}=g_{\mu\nu}g^{\sigma\rho}D^{ac}_{\sigma}D^{cb}_{\rho}+
2f^{abc}F_{\mu\nu}^{c}+R_{\mu\nu}\delta_{ab}
\end{equation}
where
\begin{displaymath}
D_{\mu}^{ac}B_{\nu}^{c}=\partial_{\mu}B_{\nu}^{a}+
\Gamma_{\mu\nu}^{\sigma}
B_{\sigma}^{a}+f^{abc}A_{\mu}^{b}B_{\nu}^{c}
\end{displaymath}
We obtain a path integral representation as in secs.2-3
 \begin{equation}
 \begin{array}{l}
W_{q}=\frac{1}{2}Tr\ln {\cal M}=-\frac{1}{2} \int dx\int_{0}^{\infty} \frac{d\tau}{\tau}
   \tau^{-\frac{2}{1+\gamma}}
  \cr
   \langle E[ \delta(q_{\tau})
 Tr({\cal T}(\tau,F))]\rangle
\end{array}
\end{equation}
where ${\cal T}$ is the solution of the equation
\begin{equation}
d {\cal T}_{ap;cr}=
A_{\mu}^{m}f_{psm}{\cal T}_{as;cr}dq^{\mu}+
\omega_{\mu a}^{b}{\cal T}_{bp;cr}
dq^{\mu}+\frac{1}{2}R_{ab}{\cal T}_{bp;cr}d\tau +
f_{psm}F^{m}_{ab}{\cal T}_{bs;cr}d\tau
\end{equation}
In order to perform the integral over the gravitational field in
eq.(29) we  need some rough approximations. First,  in the
stochastic representation of the trace (29) we apply the same
method as in eq.(22) writing $q_{s}=x+\tau^{\sigma}(\tilde{q}_{s}-x)$.
Then, for small $\tau$ we may assume that $F(q_{s})$ depends only
on $x$. Next, we  assume that only one component of $F$ is
different from zero (e.g. $ f_{psm}F^{m}=\epsilon_{ps3}F^{3} $ ).
There remains the main difficulty to  take an average of ${\cal
T}$ over the gravitational field. In these computations we make
the approximation $\omega\simeq  \langle \omega\rangle
 \simeq 0$ , $q_{\mu}\simeq b_{\mu}$ and  set
$R_{\mu\nu}\simeq \langle R_{\mu\nu}\rangle=Q\delta_{\mu\nu}$.
 Then, we can calculate $W_{q}$ exactly  \cite{habaprd}
\begin{equation}
\begin{array}{l}
 W_{q}=-\frac{1}{2} \int dx v(x)\int_{0}^{\infty} \frac{d\tau}{\tau}
   \tau^{-\frac{2}{1+\gamma}}
 \cr
  \Big(\exp(-\frac{\tau}{2}Q)\frac{u_{+}\tau}{\sinh u_{+}\tau}\frac{u_{-}\tau}{\sinh u_{-}\tau}
  \cosh (u_{+}\tau+u_{-}\tau)\cosh(u_{+}\tau-u_{-}\tau) -1 +\frac{\tau}{2}Q \Big)
\end{array}
\end{equation}
where
\begin{equation}
u_{\pm}=\frac{1}{4\sqrt{2}}\Big((F_{\mu\nu}F_{\mu\nu}+
  F_{\mu\nu}^{*}F_{\mu\nu})^{\frac{1}{2}}\pm
                   ( F_{\mu\nu}F_{\mu\nu}-F_{\mu\nu}^{*}F_{\mu\nu})
                   ^{\frac{1}{2}}\Big)
                   \end{equation}
 It follows from the behavior of the
 integrand (33) for a small $\tau$ (assuming that
 the integral (33)  is convergent for a large $\tau$) that
 for large $F^{2}$
 \begin{equation}
 W_{q}\approx F^{\frac{2}{1+\gamma}}
 \end{equation}
 In the  tree approximation we have to
 perform the Legendre transform in order
 to obtain the effective action from the
 generating functional (see the discussion at
 the scalar field).
 In the Abelian case we would obtain
  \begin{equation}
 \Gamma_{tree}(A)\simeq \frac{1}{4}\int A(\langle {\cal M}^{-1}\rangle)^{-1}A
  \end{equation}
  Note that the behavior (5) for the $\langle A A\rangle$ correlations
  means the behavior $\sim \vert k\vert^{-2\gamma}$  for the
  $\langle F(k)F(k)\rangle$ correlations (the spectral function).
  This can lead to the simplest experimental check of
  the predictions of quantum gravity if the photon
  spectral density would be measured in astrophysical observations.
  Summarizing,  for these values of the fields
  where the $\tau$ integral in eq.(33) is convergent
  the effective action consists  of the tree part
  which grows quadratically for large fields $F$,
  and the 1-loop part with slower growth (35).

    On the basis
    of the generating functional (24) we can give an argument
    for  a finiteness of the electromagnetic energy. From eq.(5) in the static approximation
it follows that the potential between  point charges is $V(r)\simeq
 r^{-1+2\gamma}$. Hence, the electric energy density
 $\epsilon \simeq r^{-4+4\gamma}$ . The
 electromagnetic  energy of the
 point charge (which includes the interaction
  energy between the charge and the gravitons) will be finite if $\gamma >\frac{1}{4}$.

\end{document}